\expandafter\ifx\csname natexlab\endcsname\relax\fi

\documentclass[12pt,preprint]{aastex}
\def\apjs{{\rm ApJS}}                  
\def\apjl{{\rm ApJ Let}}               

\def\Msol{\thinspace\hbox{$\hbox{M}_{\odot}$}}

\def\a4{\hsize 17.0cm \vsize 25.cm}

\newcommand{\dif}     {{\rm d}}

\shorttitle{clusters}

\begin{document}
\title{\bf On the star formation efficiencies and  evolution of  multiple stellar generations in 
Globular Clusters}

\author{Guillermo Tenorio-Tagle\altaffilmark{1},   
 Sergiy Silich\altaffilmark{1}, Jan Palou\v s\altaffilmark{2}, Casiana Mu\~noz-Tu\~n\'on\altaffilmark{3},
Richard W\"unsch\altaffilmark{2} }

\altaffiltext{1}{Instituto Nacional de Astrof\'\i sica \'Optica y
Electr\'onica, AP 51, 72000 Puebla, M\'exico; gtt@inaoep.mx}

\altaffiltext{2}{Astronomical Institute, Czech Academy of Sciences, Bo\v{c}n\'\i\ II 1401, 141 00 Prague, Czech Republic}

\altaffiltext{3}{Instituto de Astrof\'\i sica de Canarias
cmt@iac.es.}

\begin{abstract}
By adopting empirical estimates of the Helium enhancement ($\Delta Y$) between consecutive stellar generations for  a sample of Galactic globular clusters (GGC), we uniquely constraint the star formation efficiency ($\epsilon$) of each stellar 
generation in these stellar systems. 
In our approach,  the star formation efficiency  ($\epsilon$) is the central factor that links stellar 
generations as it  defines both their stellar mass and the remaining mass available for further star 
formation, fixing also the amount of matter required to contaminate the next stellar generation. In this 
way, $\epsilon$ is here shown to be fully defined by the He enhancement between successive stellar 
generations in a GC.  

Our approach has also an impact on the evolution of clusters and thus considers the possible loss of stars through evaporation, tidal interactions and stellar evolution.  We focus on  the  present mass ratio between consecutive stellar generations ($M_{(j-1)G}/M_{(j)G}$) and the present total  mass of Galactic globular clusters ($M_{GC}$). Such considerations suffice to determine the relative proportion of stars of consecutive generations that remain today in globular clusters ($\alpha_{(j-1)G}/\alpha_{(j)G}$). 
The latter is also shown to directly depend on the values of $\Delta Y$ and thus the He enhancement between consecutive stellar generations in GGC places  major constraints on models of star formation and evolution of GC.

\end{abstract}


\section{Introduction}
\label{intro}

Recent spectroscopic and photometric observations of Galactic globular clusters (GGCs) have led to  
the unexpected discovery of multiple stellar generations in these stellar systems.  One can indeed find a 
stellar population,  {\sl perhaps the first population}, with chemical abundance ratios similar to those 
found in neighbouring field halo stars, and within the same volume  additional  populations, characterised
by their own specific abundances of He, C, N, O, Na, Mg and Al, different to  field halo stars.
As stated by \citet{Cassisi2017}: "these abundance patterns give origin, within individual clusters, to 
well defined anti-correlations between pairs of light elements, the most characteristic one being the 
Na-O anti-correlation, nowadays considered a prominent signature for the presence of multiple 
populations in a given GGC". Similarly, multiple populations can be separated into distinct sequences in 
various color magnitude diagrams (CMD). Good examples of such photometric results, indicating 
discrete generations, have been  derived for  the two generations in NGC 6266 \citep{Milone2015A}
and in NGC 6397  \citep{Milone2012A}. As well as for the three generations of NGC 6752 
\citep{Milone2013} and the four generations of NGC 2808  \citep{Marino2014,Milone2015B,
DAntona2016}. For all of these clusters in our sample, definite results (He enhancement values, chemical 
composition, etc.) are given for each of  the identified populations. 

There is a general consent that the  light element variations are  produced by high temperature 
proton captures either in asymptotic giant branch (AGB) stars  \citep[e.g.][]{DAntona2002},  or 
in fast rotating massive stars \citep[FRMSs, e.g.][]{Prantzos2006}, or in massive interacting 
binaries \citep[see][]{DeMink2009, Schneider2014}, or red supergiant stars 
\citep{Szecsi2018}, or supermassive stars \citep[e.g.][]{Denissenkov2014, 
Denissenkov2015},  or very massive stars \citep{Vink2018}. The processed matter is then 
injected into the ambient gas by stellar winds, being able then to mix with either recently accreted 
primordial gas or with the matter left over from star formation.

All the suggested scenarios for the formation of multiple stellar populations in GGCs, as well as the 
mentioned possible polluters are affected by significant drawbacks (we refer the reader to the exhaustive 
reviews by \citet{Renzini2015,Bastian2018}. However, despite the existing
limitations, here we explore how the distinct sub-populations formed in each individual cluster with their 
own chemical peculiarities. 

Two very important constraints - obtained by an accurate analysis of suitable photometric 
datasets are the helium abundance difference between distinct sub-populations and their present 
relative population ratios \citep{Cassisi2017,Milone2018}. This is because the various stellar 
candidate polluters predict different amounts of He enhancement in quite distinct evolutionary phases, 
and the analysis of present He ratios between the first stellar generation and the second or more, 
poses a critical constraint on scenarios aimed to explain the origin of the multiple population phenomenon 
\citep[see, e.g.,][]{Bastian2015A,TenorioTagle2016}.

In a previous paper \citep[][here after Paper I]{TenorioTagle2016}
by adopting the empirical estimates of He enhancement ($\Delta{Y}$), present mass ratio between first 
and second stellar generations ($M_{1G}/M_{2G}$) and the actual mass of GGCs ($M_{GC}$), we 
constraint  the star formation efficiency of  Galactic GCs. The model was limited by considering only two 
stellar generations in each individual cluster, when at that time there were few clusters showing evidence 
of triple (or more) stellar sub-populations such as the cases of NGC 2808 \citep{Milone2015A} and 
NGC 6752 \citep{Milone2013}. 
Nowaday, the number of GGCs which have been proved to host more than two stellar populations is 
largely increased mainly by the photometric dataset collected in the framework of  the {\it Hubble Space 
Telescope UV Legacy Survey of Galactic GCs} project \citep{Milone2018}. This project has also 
allowed for the homogeneous determination of $\Delta {Y}$ for a large sample of clusters.

Here, we wish to extend the analysis performed in Paper I to a larger sample of clusters,  hosting more than two stellar generations.

The plan of the paper is as follows: Section 2 describes all major  assumptions in the model. Section 3 shows the direct dependance of  $\Delta {Y}$ on the efficiency of star formation for each stellar generation.
Section 4 shows how the values of $\Delta{Y}$ between consecutive generations, impact on the 
corresponding ratio of the number of stars that remain today as part of the cluster. Section 5 derives,
under some assumptions, the total mass in stars formed in each generation as well as the mass of the clouds that originate them. Section 6 presents our main conclusions and an evaluation of our model.

\section{Main assumptions}

Here we briefly outline the main assumptions at the basis of our considerations.
We assume a massive pristine cloud that has been uniformly contaminated with the products of supernovae from the first stars in the galaxy, population III stars, and thus presents a uniform Fe abundance. This is to be acquired by all stellar generations produced by the cloud, unless it manages to capture the products from subsequent  supernovae enhancing then the Fe abundance of the remaining cloud and of further stellar generations. This is to cause the so called "Fe spread", as detected among the various stellar generations belonging to the most massive globular clusters.  

The cloud with a total mass $M_{tot}$,  collapses to form multiple stellar generations, each of these with 
a peculiar chemical pattern resulting from  the  contamination of the residual gas by H burning products 
cooked at high temperatures by the former stellar generation.  It is assumed that the negative stellar feedback in dense compact clusters is drastically suppresed   
\citep[see][]{Silich2017,Silich2018} and leads to completely opposite results to those found in the adiabatic cluster wind model of \citet{Chevalier1985}
in which all the gas left over from the star formation process is not considered as it is assumed to have 
been totally dispersed by the feedback caused by massive stars. Following  Paper I, we assume that all 
stellar generations -- at their formation -- sample a full 
\citet{Kroupa2001A} initial mass function (IMF) with stars in the range 0.1 - 120 M$_\odot$ and, as 
expected from massive  starbursts restricted to a small volume, a large fraction of their massive stars are 
to end up as interacting binaries \citep[see][]{DeMink2009, Schneider2014} contaminating the left over 
gas. The contamination comes from the large fraction (about 70 $\%$) of massive star that 
are in binaries and that will interact at some time during their evolution, consistent with 
observations of local high-mass star forming regions \citep{Bastian2013,Sana2012,Izzard2013}.

 As in Paper I,  massive binaries are here held responsible of contaminating the left over gas with H burning products while effectively holding the further collapse of the cloud. It is not until the end of the type II supernova (SN) era (say $\sim$ 40 Myr after the onset of the last episode of star formation) that gravity 
wins again and the collapse of the left over cloud, now contaminated with the products of a former 
generation,  leads to another episode of star formation \citep[see also][]{Kim2018}. With these 
assumptions  in hand, one can 
envisage numerous episodes of star formation and further contamination of the residual gas until the latter is either exhausted (turned into stars) or dispersed by (feedback) the mechanical energy injected into the cluster volume by the last stellar generation.  
Indeed, SN have been declared responsible by several authors of dispersing the gas left over 
after star formation \citep[see, for example][]{Calura2015}. This inevitably leads to the 
assumption of a later accretion as in the model of \citet{DErcole2008}, to ensure the sufficient 
matter to form a second stellar generation and avoid a mass budget problem. Other scenarios as 
the dark remnant accretion scenario of \citet{Krause2012,Krause2013} find a limiting initial 
cloud mass of $10^7$\Msol \, above which the gas may not be ejected and therefore form 
additional stars. In this respect \citet{Silich2017,Silich2018} defined the physical conditions that
lead to dispersion or retention of the parental cloud. 

In our scenario the blowout of individual SN and the fragmentation of their shells of swept up matter, are 
also central to our considerations. 
SN explosions  lead to a highly supersonic shock wave that sweeps into an expanding shell all the matter 
that it encounters. All of it, whether it is the nearby wind of the secondary or the normal ISM or the gas 
left over from star formation. As this happens the kinetic energy of the ejected matter is being converted 
into thermal energy, after crossing the inner or reverse shock. The remnant structure presents then an 
outer expanding shell that contains all the swept up matter and that continuously acquires the new 
material overtaken by the leading shock. The back of the shell is bound by a Contact Discontinuity 
that encloses all of the thermalised, ejecta. As both the hot ejecta and the outer shell
expand with the same speed, there is not mixing at all between the swept up gas and the SN 
products.
    
Sooner or later however,  the supernova remnant will find the edge of the cloud and this will lead 
to its blowout. As a section of the leading shock finds the density gradient it will speed up radially and 
away from the cloud center. The same will happen to the corresponding section of the shell behind the 
accelerated shock. However, its sudden acceleration will drive the shell Rayleigh-Taylor unstable and this 
will lead to its fragmentation. The hot thermalised ejecta will then rapidly move between the shell 
fragments and follow the leading shock out of the cluster, while the remains of the shell, with most of 
the swept up matter, having no longer a driving pressure ceases to expand and easily mix with the left 
over cloud  \citep[see][]{TenorioTagle2015,TenorioTagle2016}.      

If one demands a full IMF, we believe the blowout of SN is the key to understand 
why most GCs show 
the same Fe metallicity in their stellar generations. Otherwise, the retention of SN products should cause 
a metallicity spread between generations, fact that only seems to have occurred in very small amounts 
(less than 5$\%$) in the most massive GCs, which clearly were able to capture a SN remnant.
 
Blowout of SN is also expected if the SN shell is able to overtake several of the neighboring wind sources, 
as this enhances the thermal  energy content causing its acceleration and sudden fragmentation even if 
evolving into a constant density medium \citep[see][]{Silich2017}. In this way the ejecta 
from SN is to be eventually expelled out of the cluster without changing the metallicity of the leftover 
gas. Only SN exploding in the central and denser regions of the most massive and compact clouds
are expected to become  pressure confined  and then be able to eventually cause an Fe contamination of 
the residual gas.     

As in Paper I, we base our analysis on the measured helium abundance enhancement ($\Delta Y$), and 
the present mass ratio between consecutive stellar generations, but no attempt was made to correlate 
these empirical constraints with other observed peculiarities such as, for instance, the light elements 
abundance: which are out of the scope of our analysis. Note however that to determine the empirical 
values of  $\Delta Y$ that best reproduces the observations, a fine grid of synthetic spectra, based on the 
average abundance of at least 22 different elements in each of the detected stellar generations   is used. 
The absolute  value of the synthetic color distance  between two stellar generations at each of  the 
observed filters minus the corresponding observed color distance between the two stellar generations 
leads to the best  $\Delta Y$ value 
\citep{Sbordone2011,Milone2012A,Milone2012B,Milone2013}.
Note also that the synthetic spectra used for  each of the stellar generations, warrants without knowledge of the possible contaminer(s), the "correct" elemental abundance of the gas out of which further stellar generations were formed.

\section{$\Delta Y$ constraints on the star formation 
efficiency of subsequent stellar generations}

After the  gravitational collapse of the pristine cloud, the mass in stars of the first  stellar 
generation is:
\begin{equation}
\label{eq:M1G}
M_{1G} = \epsilon_{1G} M_{tot} 
\end{equation}
where $\epsilon_{1G}$ is the efficiency of star formation and $M_{tot}$ is the mass of the 
pristine cloud. Here Interacting Massive Binary (IMB) stars are considered as polluters and 
therefore  we used  the output from the Binary Population and Spectral Synthesis code, 
BPASS Version 2.1 \citep[][see panel a in Figure 1]{Eldridge2017} 
to obtain the cumulative masses of the ejected hydrogen and helium: 
\begin{eqnarray}
      \label{eq2a}
      & & \hspace{-0.9cm} 
M_{H}(t) = \int_0^t {\dot M}_H \, \dif{t} ,
      \\[0.2cm] \label{eq2b}
      & & \hspace{-0.9cm} 
M_{He}(t) = \int_0^t {\dot M}_{He} \, \dif{t} .
\end{eqnarray}
\begin{figure}[htbp]
\plottwo{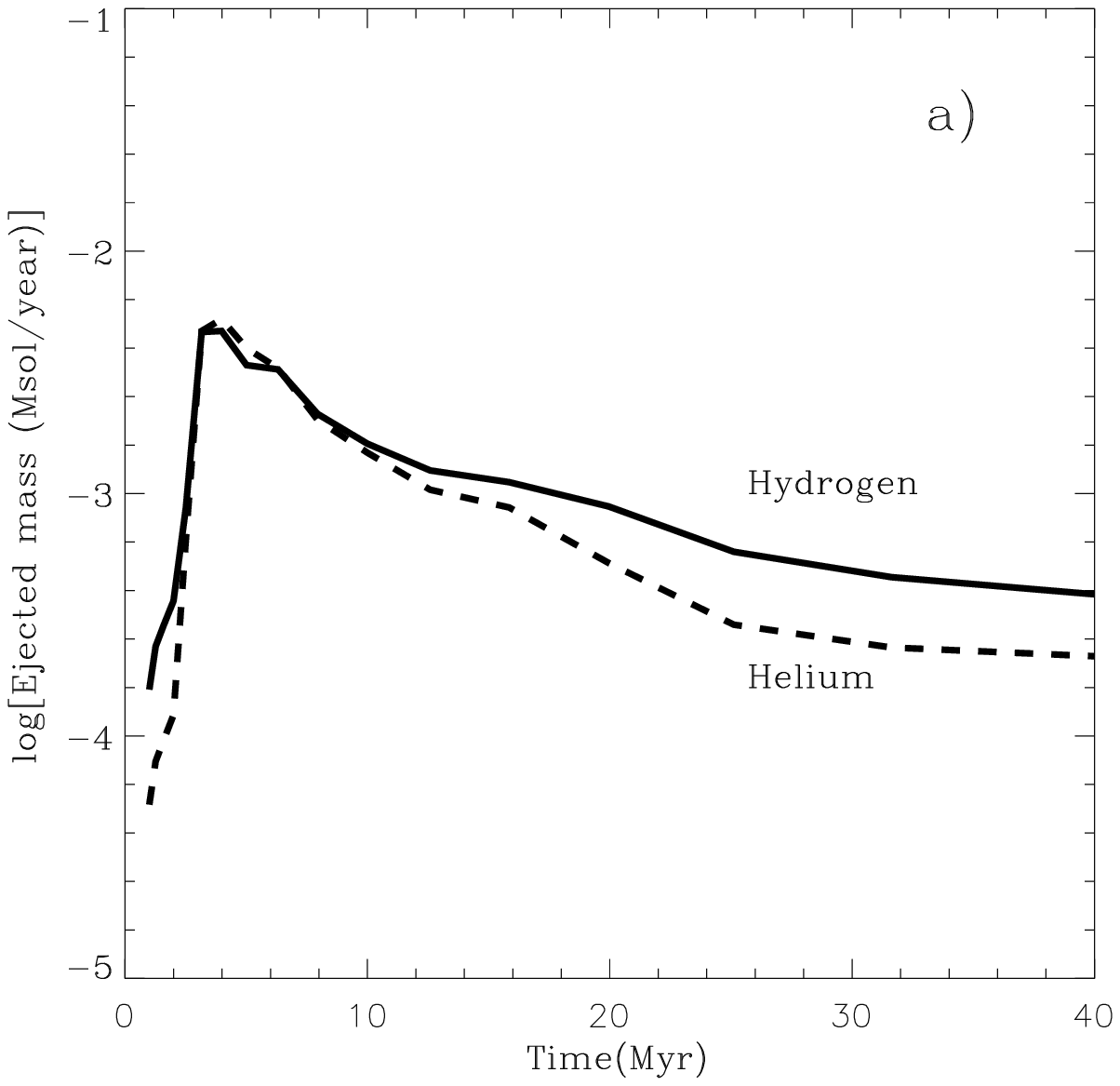}{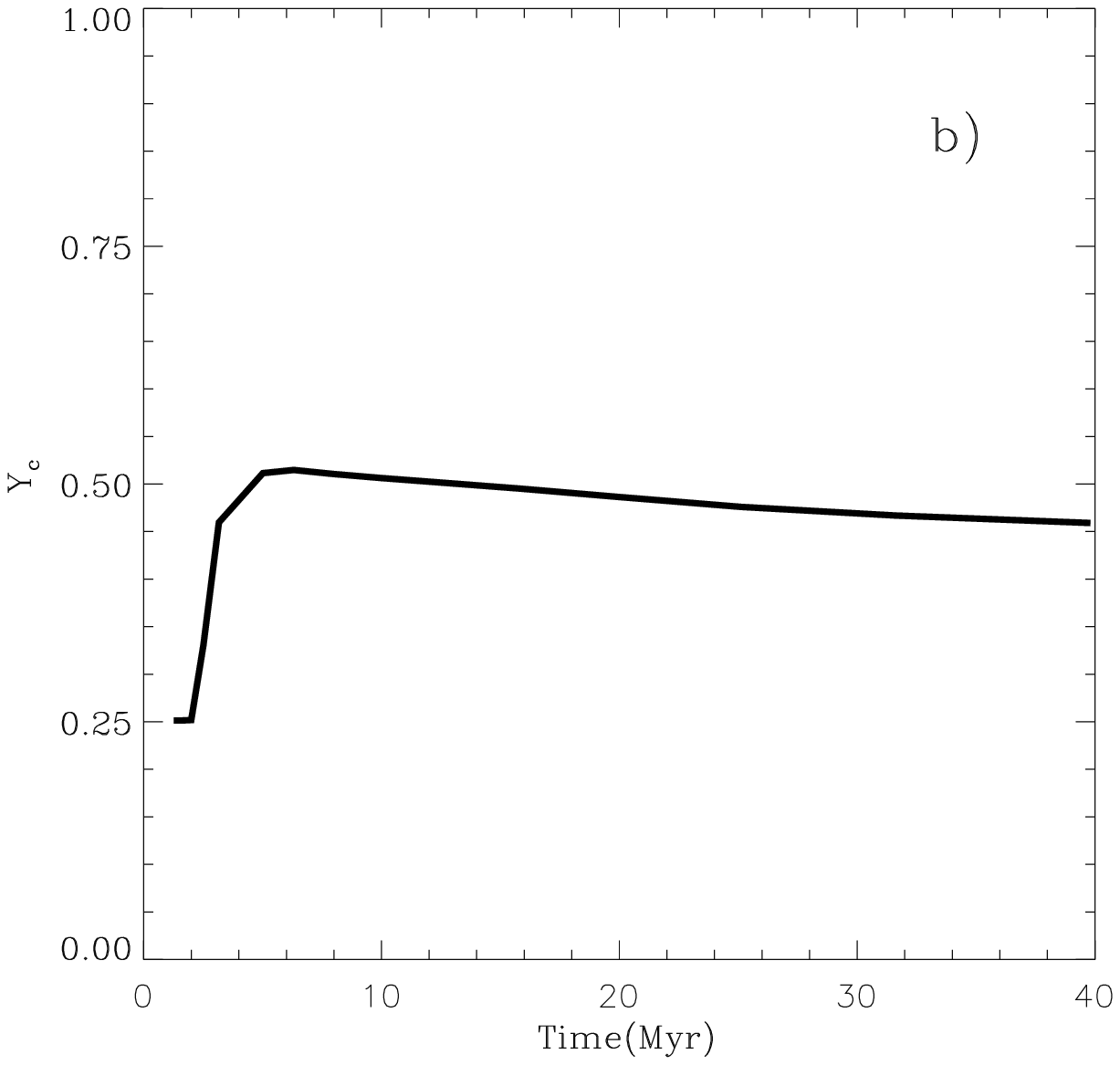}
\caption{The reinserted mass and the He mass fraction. The H and He mass loss rates predicted by the 
BPASS model (panel a) and the cumulative He mass fraction $Y_c$ (panel b) as functions of time.}
\end{figure}
This yields for a cluster formed with its full IMF in a cloud with a metallicity Z = 0.001, after 40 Myr of 
evolution, a fraction of the stellar mass shed by massive stars, including binaries, equal to: 
$M_{(c)} = 0.083 M_{1G} = 0.083  \epsilon_{1G} M_{tot}$. This is  assumed to mix thorougly with  
the mass of pristine gas left over after the 1G has formed: $M_{(p)} = (1 - \epsilon_{1G}) M_{tot}$.
In this way  the helium mass fraction of the mixed gas, $Y$, is:
\begin{equation}
\label{eq:Y}
Y = \frac {(1 - \epsilon_{1G}) Y_{(p)}  + 0.083 \epsilon_{1G} Y_{(c)} }   {1 - 0.917 \epsilon_{1G}} ,
\end{equation}
where the cumulative He mass fraction $Y_c$ (see Figure 1, panel b) is:
\begin{equation}
      \label{eq1b}
Y_c (t) = \frac{M_{He}(t)}{M_H(t) + M_{He}(t)} . 
\end{equation}
After 40 Myr of evolution the He mass fraction for our low metallicity cluster  is $Y_{(c)}$ = 0.459, 
while the primordial value $Y_{(p)}$ = 0.245.
The corresponding value of the helium enhancement ($\Delta Y$) in the left over gas then is:
\begin{equation}
\label{eq:DY}
\Delta Y = Y - Y_{(p)} = \frac {0.083 \epsilon_{1G} (Y_{(c)} - Y_{(p)})} {1- 0.917 \epsilon_{1G}}
\end{equation}
Following  the above assumptions one can derive the values of the Helium enhancements $\Delta Y_{(j-1, j)}$ for subsequent stellar generations: 

\begin{equation}
\label{eq:DYjm1j}
\Delta Y_{(j-1, j)} = Y_j - Y_{(j-1)} = \frac {0.083 \epsilon_{(j-1)G} (Y_{(c)} - Y_{(j-1)})} {1- 0.917 \epsilon_{(j-1)G}}
\end{equation}

Thus, as claimed in paper I, the efficiency of star formation of the previous stellar generation ($\epsilon_{(j-1)G}$) also defines both the total amount of gas left over from star formation as well as the mass of the contaminer gas  ($0.083 M_{(j-1)G}$) which  leads, upon a thorough mixing, to a contaminated cloud ready to trigger another stellar generation with its own He abundance $Y_{(j)}= Y_{(j-1)} + \Delta Y_{(j-1, j)}$. Thus,  if  $\Delta Y$ is measured between generation j and j-1. such a quantity ($\epsilon_{(j-1)G}$) is then fully defined by the observations, by  the values of $\Delta Y_{(j-1, j)}$:
\begin{equation}
\label{eq8}
\epsilon_{(j-1)G} = \frac{\Delta Y_{(j-1, j)}}  {0.083(Y_{(c)} - Y_{(j-1)}) + 0.917 \Delta Y_{( j-1, j)}}
\end{equation}
\begin{figure}[htbp]
\plotone{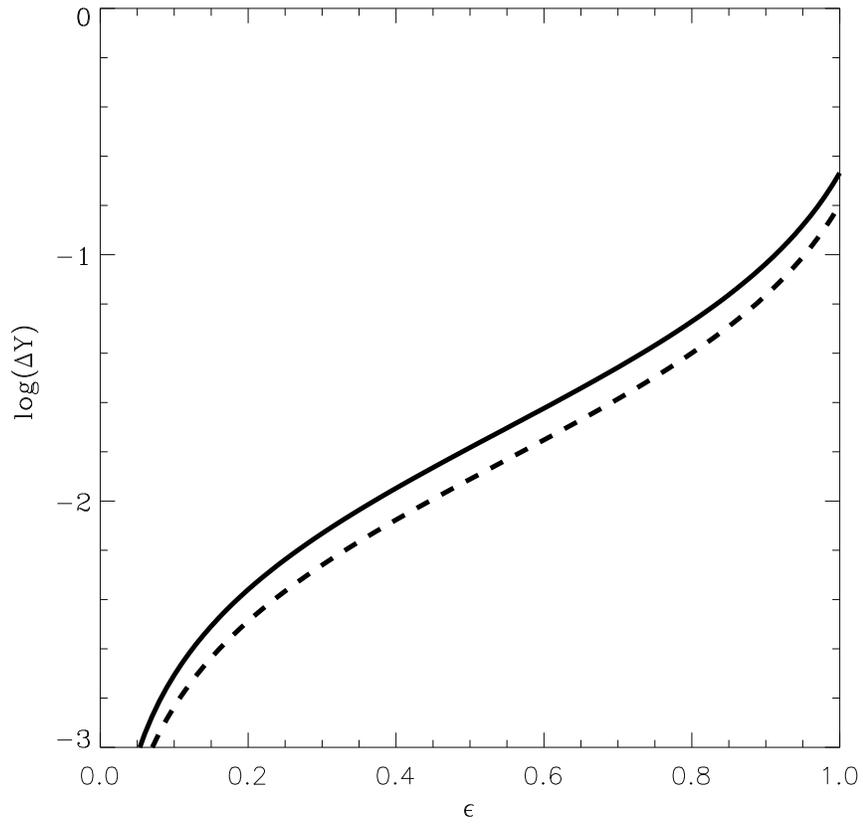}
\caption{log $\Delta Y_{(j-1, j)} {\it vs} $ the star formation efficiency ($\epsilon$). The
solid line presents the results from equation \ref{eq:DY} assuming $Y_{(j-1)} = Y_p $.
The dashed line is for $Y_{(j-1)}$ = 0.30 and shows the largest expected change in the 
location of the solid line (see text).}
\label{fig1}
\end{figure}
According to equation \ref{eq8}, the $\Delta Y_{( j-1, j)}$ values between subsequent generations lead to the efficiency of star formation of all but the last stellar generation (see Figure \ref{fig1}).  Note that the factor 
$0.083 (Y_{(c)} - Y_{(j-1)})$ shifts the location of this line. However, even if one uses an extremelly large value of $Y_{(j-1)} \sim 0.30$, as found for 
the blue main sequence in $\omega$ Centauri, which represents the largest "super-helium rich" 
population ever found in a GGCs 
\citep{Bedin2004,Salaris2004},
the changes in the resultant location of equation \ref{eq8} are marginal (see Figure \ref{fig1}). 
And thus, within this framework, cases with small values of $\Delta Y$, lead to small values of 
$\epsilon_{(j-1)G}$ and conversely,  large $\epsilon_{(j-1)G}$, result from large $\Delta Y$ values 
(see Figure \ref{fig1}).

Equation \ref{eq8} provides then a way to derive the star formation efficiencies of multiple generations in  globular clusters. This method is completely different  than the traditional one in which the star formation efficiency $\epsilon$ = mass in stars / mass of progenitor cloud. This is unapplicable to globular clusters, as their stellar mass is largely modified by many Gyrs of stellar and dynamical evolution, and the mass of the progenitor cloud is completely unknown.

\subsection{The total efficiency of star formation ($\epsilon_{tot}$)}

In our approach the total mass available for stellar generation (j)G is:

\begin{equation}
\label{eq:Mj1}
M_{(j)} = (1 - \epsilon_{(j-1)G}) M_{(j-1)} + 0.083  \epsilon_{(j-1)G} M_{(j-1)} = (1 - 0.917 \epsilon_{(j-1)G}) M_{(j-1)}
\end{equation}

\noindent where the first term accounts for the fraction of the gas left over after the formation of the (j-1) stellar generation ($(1-\epsilon_{(j-1)G})M_{(j-1)}$) and the second term for the reinserted mass that contaminates the left over gas. The only exception is  the first stellar generation for which there is no previous injection of matter and thus  the total available mass for star formation is $M_{tot}$. One can express  $M_{(j-1)}$ also as a function of the mass left from all previous generations. In such a case $M_{(j)}$ is:
\begin{equation}
\label{eq:Mj2}
M_{(j)} = (1 - 0.917 \epsilon_{(1)G}) (1 - 0.917 \epsilon_{(2)G}) \times .... \times  (1 - 0.917 \epsilon_{(j-1)G}) M_{tot}
\end{equation}
As in equation \ref{eq:M1G}, this equation multiplied by a star formation efficiency factor  
($\epsilon_{(j)G}$)  leads to the mass of the next stellar generation.

To calculate the total efficiency of star formation ($\epsilon_{tot}$) in our scheme,  requires to
take into consideration the amount of matter returned by massive stars to cause the contamination of the left over gas and then be used again in further episodes of star formation.    In such a case 
\begin{eqnarray}
      \nonumber
      & & \hspace{-1.1cm} 
M_*  = M_{tot} (\epsilon_{1G} + \epsilon_{2G}  (1 - 0.917 \epsilon_{(1)G}) + \epsilon_{3G}  (1 - 0.917 \epsilon_{(1)G}) (1 - 0.917 \epsilon_{(2)G}) + ... 
 \\[0.2cm]     \label{eq:Mstar}
      & & \hspace{-1.1cm}
+ \epsilon_{(n)G}  (1 - 0.917 \epsilon_{(1)G}) (1 - 0.917 \epsilon_{(2)G})
... (1 - 0.917 \epsilon_{(n-1)G})) = \epsilon_{tot} M_{tot} 
\end{eqnarray}
and thus  $\epsilon_{tot}$ is equal to the sum of the efficiencies of all stellar generations, taking into 
consideration however, the fraction of the remaining gas available for each of them:
\begin{eqnarray}
     \nonumber
      & & \hspace{-1.1cm} 
\epsilon_{tot}  =  \epsilon_{1G} + \epsilon_{2G}  (1 - 0.917 \epsilon_{(1)G}) + \epsilon_{3G}  (1 - 0.917 \epsilon_{(1)G}) (1 - 0.917 \epsilon_{(2)G}) + ... 
 \\[0.2cm]     \label{eq:epstot}
      & & \hspace{1.0cm}
+ \epsilon_{(n)G}  (1 - 0.917 \epsilon_{(1)G}) (1 - 0.917 \epsilon_{(2)G})
... (1 - 0.917 \epsilon_{(n-1)G})
\end{eqnarray}
Values of $\epsilon_{tot}$ for the clusters here considered are given in Table1.

\section{$\Delta Y$ constraints on the dynamical evolution of Globular Clusters}

One can also consider the  evolution of stellar clusters, accounting for the possible loss of stars through 
evaporation, tidal interactions or stellar evolution. In such a case,  equations \ref{eq:M1G} and 
\ref{eq:Mj2},  once multiplied by $\epsilon_{(j)G}$ should also be multiplied by the fraction of newly 
formed stars  that have remained until now gravitationally trapped within the cluster ($\alpha_{(j)G} 
\le$ 1) and contribute to its present mass while remaining in the cluster, $\alpha_{(j)G}M_{(j)G}$. In 
this way one can derive the mass ratio between consecutive generations and compare it with the mass 
ratio inferred from the observations ($x_{(j-1, j)} = M_{(j-1)G} / M_{(j)G}$):
\begin{equation}
\label{eq:xjm1j1}
x_{(j-1, j)} = \frac {M_{(j-1)G}} {M_{(j)G}} = \frac{ \alpha_{(j-1)G} \epsilon_{(j-1)G}}  { \alpha_{(j)G} \epsilon_{(j)G}  (1 - 0.917 \epsilon_{(j-1)G})} 
\end{equation}
This through equation \ref{eq:DYjm1j} leads to:
\begin{equation}
\label{eq:xjm1j2}
x_{(j-1, j)} = 
 \frac {\alpha_{(j-1)G} \Delta Y_{(j-1, j)} } {\alpha_{(j)G} \epsilon_{(j)G} 0.083 (Y_{(c)} - Y_{(j-1)})} 
\end{equation}
and thus 
\begin{equation}
\label{eq15}
 \frac {\alpha_{(j-1)G}} {\alpha_{(j)G}}  =   \epsilon_{(j)G} \frac {0.083 
(Y_{(c)} - Y_{(j-1)}) x_{(j-1,j)}}{\Delta Y_{(j-1, j)}} 
\end{equation}
Thus the $\alpha$ ratios are also fully constrained by the observations, for each set of consecutive 
generations. It is only when one considers the last generation (n) and the one before last (n-1), that the 
proportion would  have  to include  the still undefined $\epsilon_{(n)G}$:
\begin{equation}
\label{eq16}
\frac {\alpha_{(n-1)G}} {\alpha_{(n)G} } =    \epsilon_{(n)G} \frac {0.083 (Y_{(c)} - Y_{(n-1)}) x_{(n-1,n)}}   
{\Delta Y_{(n-1, n)}} 
\end{equation}

The $\alpha$ variables for each generations as well as the star formation efficiency for the last stellar generation ($\epsilon_{(n)G}$) cause a degeneracy in equations \ref{eq15} and \ref{eq16}. However, the proportion of $\alpha_{(j-1)} / \alpha_{(j)} $ found through equation \ref{eq15} for each pair of consecutive generations remains unchanged 
regardless of the mass assigned either to the primordial clouds
or the total stellar mass of each stellar generation.

In Table~1 we list the observed and derived parameters for some well studied  GGCs. The table indicates 
first the NGC identification number, followed by the observed total mass and the derived total efficiency 
($\epsilon_{tot}$) and the calculated range of $M_\mathrm{tot}$ (see Section 5).
The following lines provide data and derived parameters for different stellar generations in each cluster.
Columns 2-4 give  the observed mass in each generation,  the value of $Y_{j-1}$ and   the observed He 
enhancement $\Delta Y$ between generations. The last two columns 
list the efficiency of star formation, as derived from equation \ref{eq8}, and the assumed or 
derived values of the $\alpha$ ratios for each set of consecutive generations.  
\begin{table}[htp]
\caption{\label{tab:1} Observed and derived parameters for a sample of GGCs}
\begin{tabular}{l c c c c c c c c}
\hline\hline\hline
NGC & $M_{GC}/M_{\odot}$ & $\epsilon_{tot}$ &  M$_{tot}/M_{\odot}$ &   &  & &   &   \\
6266  & $ 7.07 \times 10^5$  &  0.87 + $ 0.20  \epsilon_2$ &  
$2 \times 10^6 - 2 \times 10^8$ & &  &  &  &  \\
\hline
Generation  & $M_{(j)G}/M_\odot$ &  $Y_{j-1}$ & $\Delta Y$ & $\epsilon_{(j)G}$ &   $\alpha_{(j-1)G}/\alpha_{(j)G}$ &  &     \\
\hline
$1^{st}$ &  5.585 $ \times 10^5$ & $Y_p$ = 0.245 & -  &   0.87 &   - & &  \\
2$^{nd} $   & 1.485 $\times 10^5$ & Y=0.325  & 0.08 &  -  & 0.835 $\epsilon_2$  &  &  \\

\hline\hline\hline
NGC & $M_{GC}/M_{\odot}$ & $\epsilon_{tot}$ &  M$_{tot}/M_{\odot}$ &   &  & &   &   \\
 6397  & $ 8.89 \times 10^4$ &  0.37 + $ 0.34 \epsilon_2$ &   
$2.5 \times 10^5 - 2.5 \times 10^7$ &  &     &  &   &   \\\hline
Generation  & $M_{(j)G}/M_{\odot}$ &  $Y_{j-1}$ & $\Delta Y$ & $\epsilon_{(j)G}$ &   $\alpha_{(j-1)G}/\alpha_{(j)G}$ &  &  \\
\hline
$1^{st}$ &  $2.667 \times 10^4 $ &$Y_p$ = 0.245 & -  &   0.37 &   - & &  \\
2$^{nd} $   & 6.223 $ \times 10^4$ & Y=0.255  & 0.01 &  -  & 0.76 $ \epsilon_2$  &  &  \\

\hline\hline\hline
NGC & $M_{GC}/M_{\odot}$ & $\epsilon_{tot}$ &  M$_{tot}/M_{\odot}$ &   &  & &   &   \\
 6752  & $ 2.39 \times 10^5$ &  0.729  + $ 0.33  \epsilon_3$ &  
$6.9 \times 10^5 - 5.7 \times 10^7$ & &     &  &   &  
 \\\hline
Generation  & $M_{(j)G}/M_{\odot}$ &  $Y_{j-1}$ & $\Delta Y$ & $\epsilon_{(j)G}$ &   $\alpha_{(j-1)G}/\alpha_{(j)G}$ &  &   \\
\hline
$1^{st}$ &  5.975 $\times 10^4 $ &$Y_p$ = 0.245 & -  &   0.32 &   - & &  \\
2$^{nd} $   & 1.076 $\times 10^5 $ & Y=0.254  & 0.008 & 0.58  & 0.711&  &  \\
$3^{rd}$     &  7.170 $\times 10^4 $ & Y=0.275  &  $ 0.021$ &   $ - $ & $ 1.209 \epsilon_3$ &   &  \\

\hline\hline\hline
NGC & $M_{GC}/M_{\odot}$ & $\epsilon_{tot}$ & M$_{tot}/M_{\odot}$ &   &  & &   &   \\
 2808  & $ 7.42 \times 10^5$ &  1.277  + $0.016 \epsilon_4$&  
$2.2 \times 10^7 - 2.2  \times 10^{9}$ & &  &  &  &  
 \\\hline
Generation  & $M_{(j)G}/M_\odot$ &  $Y_{j-1}$ & $\Delta Y$ & $\epsilon_{(j)G}$ &   $\alpha_{(j-1)G}/\alpha_{(j)G}$ &  &   \\
\hline
$1^{st}$ &  4.304$ \times 10^4$ & $Y_p$ = 0.245 & -  &   0.66 &  - & &  \\
2$^{nd} $   & 3.250$ \times 10^5$ & Y=0.276  & 0.03 & 0.85 & 0.066 &  &  \\
$3^{rd}$     &  2.322 $\times 10^5$  & Y=0.336  &  $ 0.06 $ &   $  0.89 $ & $ 0.3599 $ &   &  \\
$4^{th}$    & 1.417 $\times 10^5$ & Y=0.386  &  $ 0.05 $ &   $ - $   & $0.331 \epsilon_4$ &   &  \\

\hline\hline\hline
\end{tabular}
\end{table}

Please note that in NGC 6266 $M_{1G}$ is larger than $M_{2G}$ 
\citep{Milone2015A} and the opposite is true for NGC 6397 \citep{Milone2012C}.
In fact, the fraction of 1G stars ranges from about 8 per cent to 67 per cent and as shown by 
\citet{Milone2017}. \citet{Marino2014} found for a sample of 50 clusters, that the fraction of 1G stars 
anticorrelates with the present mass of the cluster.  

NGC 6752 seem to have had very low abundances in the first and second generation \citep{Milone2013}.

In NGC 2808 the total eficiency of star formation is not larger but close to 1 (see equation 
\ref{eq:epstot}) \citep{Milone2015B,DAntona2016}.
According to the last reference there are 5 stellar generations A, B, C, D and E. However, there is no 
$\Delta Y$ enhancement between generations B and C and thus we have taken these B + C as one 
single generation. This was also justified by the less evident observed separation between the main 
sequences of B and C.

\section{Estimates of the original mass of the cloud ($M_{tot}$) and of the stellar generations}

Assuming a certain value of the retained stars for the last generation, ($\alpha_{(n)G}$) and of the star 
formation efficiency of the last generation ($\epsilon_{(n)G}$) one can calculate from fractions
$\alpha_{(j-1)G}/\alpha_{(j)G}$ (equations \ref{eq15} and \ref{eq16}), absolute values of 
$\alpha_{(j)G}$ for all
previous generations. For example, in the case of clusters with  only two stellar generations 
 the values of $\alpha_{(1)G}$ for the first stellar generation are function of the selected efficiency of the last stellar generation $\epsilon_{(2)G}$ and the value selected for  $\alpha_{(2)G}$. Similarly, for clusters with more than two stellar generations one can derive 
values of $\alpha_{(n-1)G}$ for selected values of  $\epsilon_{(n)G}$ and $\alpha_{(n)G}$ and in a similar way 
use these results  to obtain values of $\alpha_{(n-2)G}$ and so on, until one obtains values of $\alpha_{(1)G}$. In all cases 
one should only keep the resultant $\alpha$ values that are $\leq$ 1. Note then that values of $\alpha_{(1)G}$  
for the first stellar generation depend directly on the selected values for the efficiency of the last stellar generation ($\epsilon_{(n)G}$) as well as on $\alpha_{(n)G}$ (see Figure 3).
  
\begin{figure}[htbp]
\vspace{18.0cm}
\includegraphics{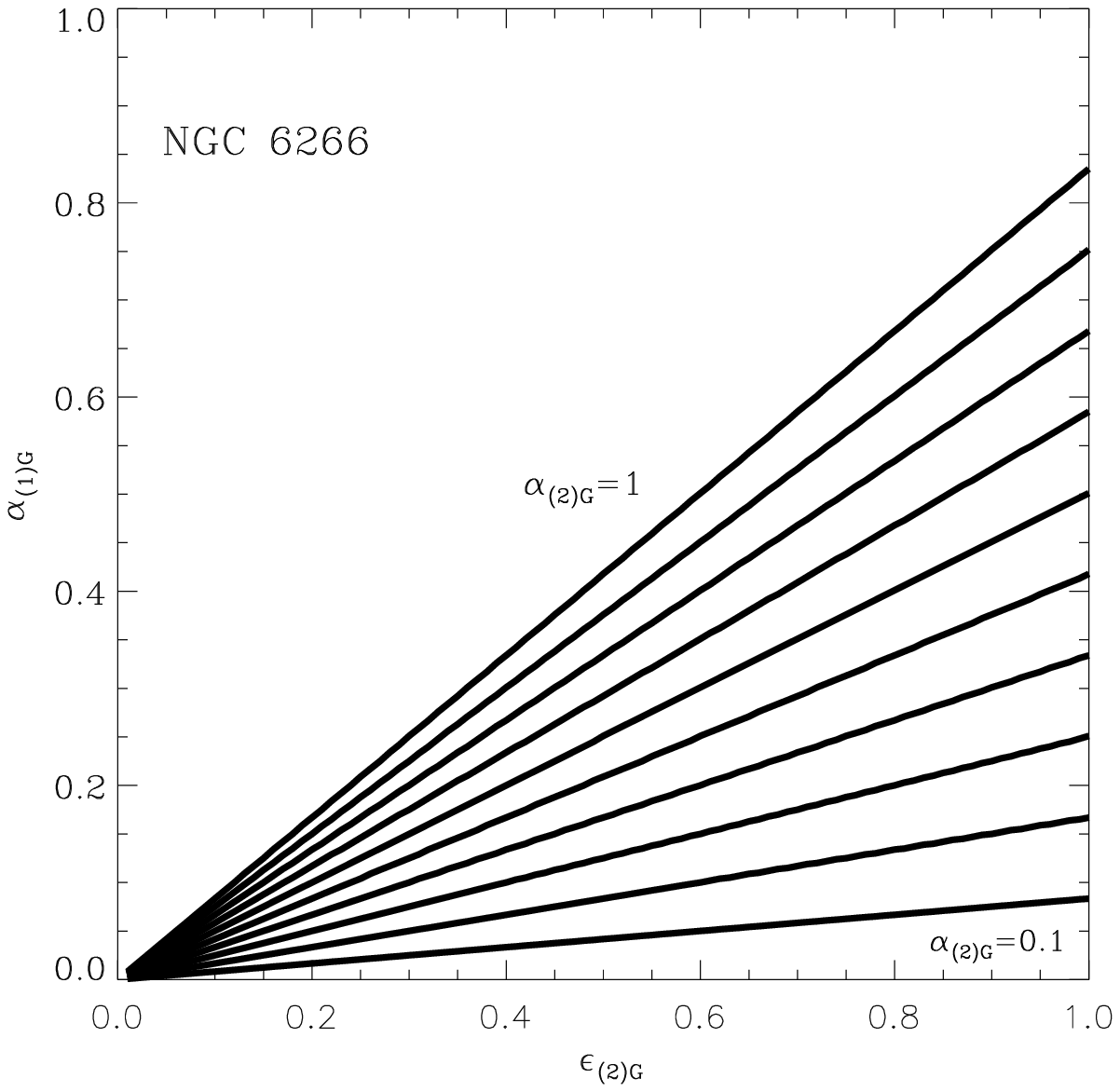}
\includegraphics{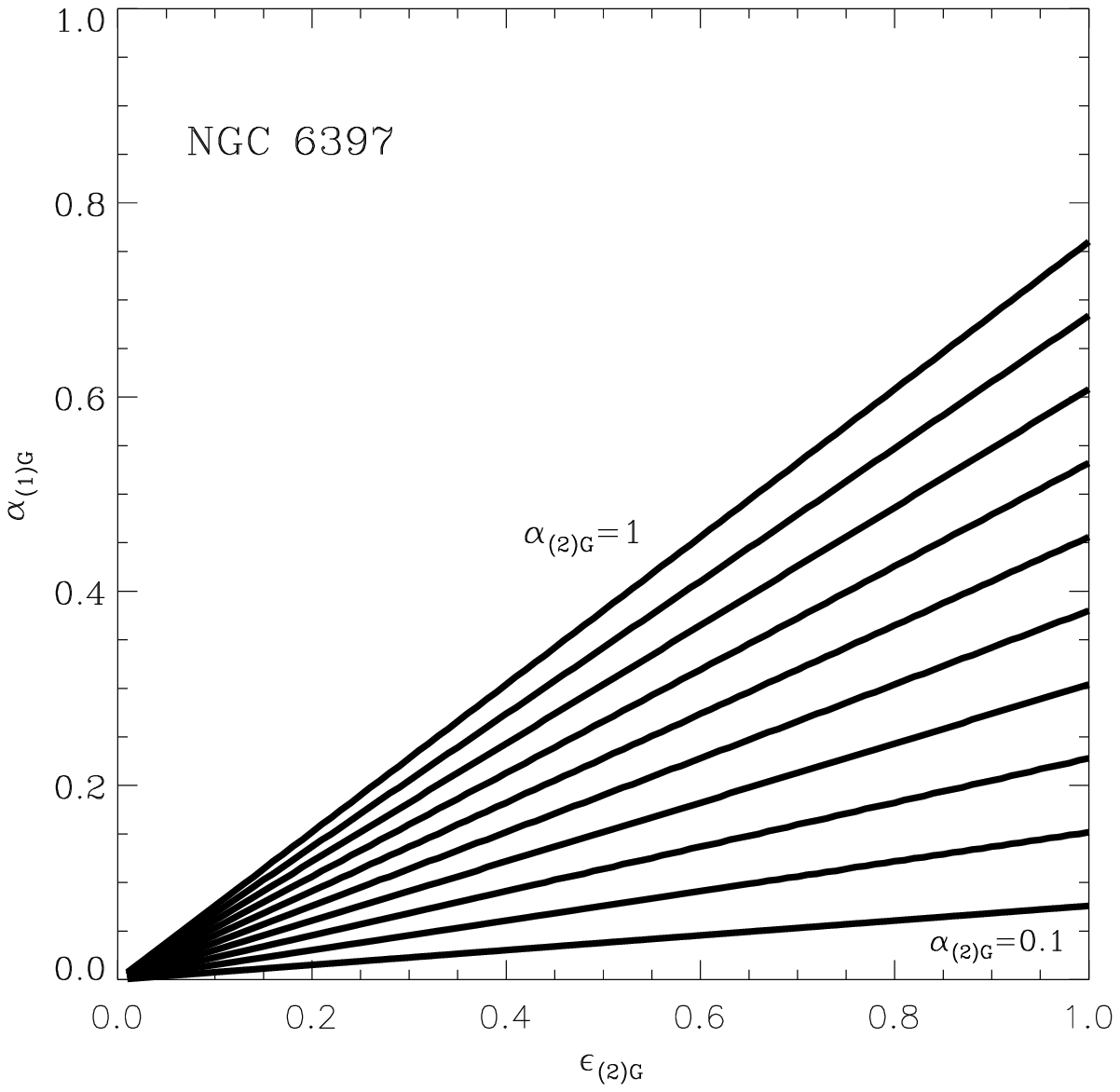}
\includegraphics{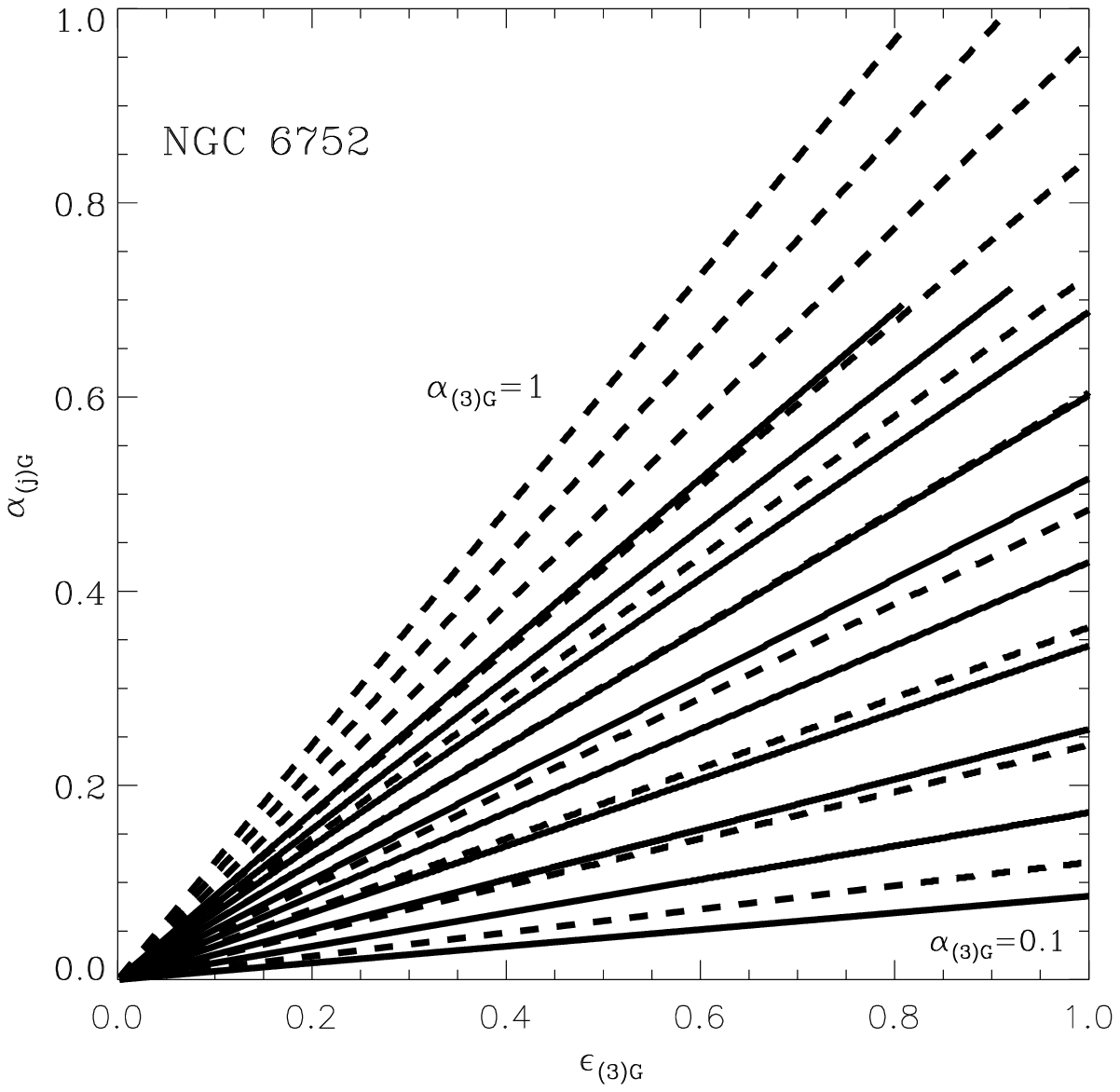}
\includegraphics{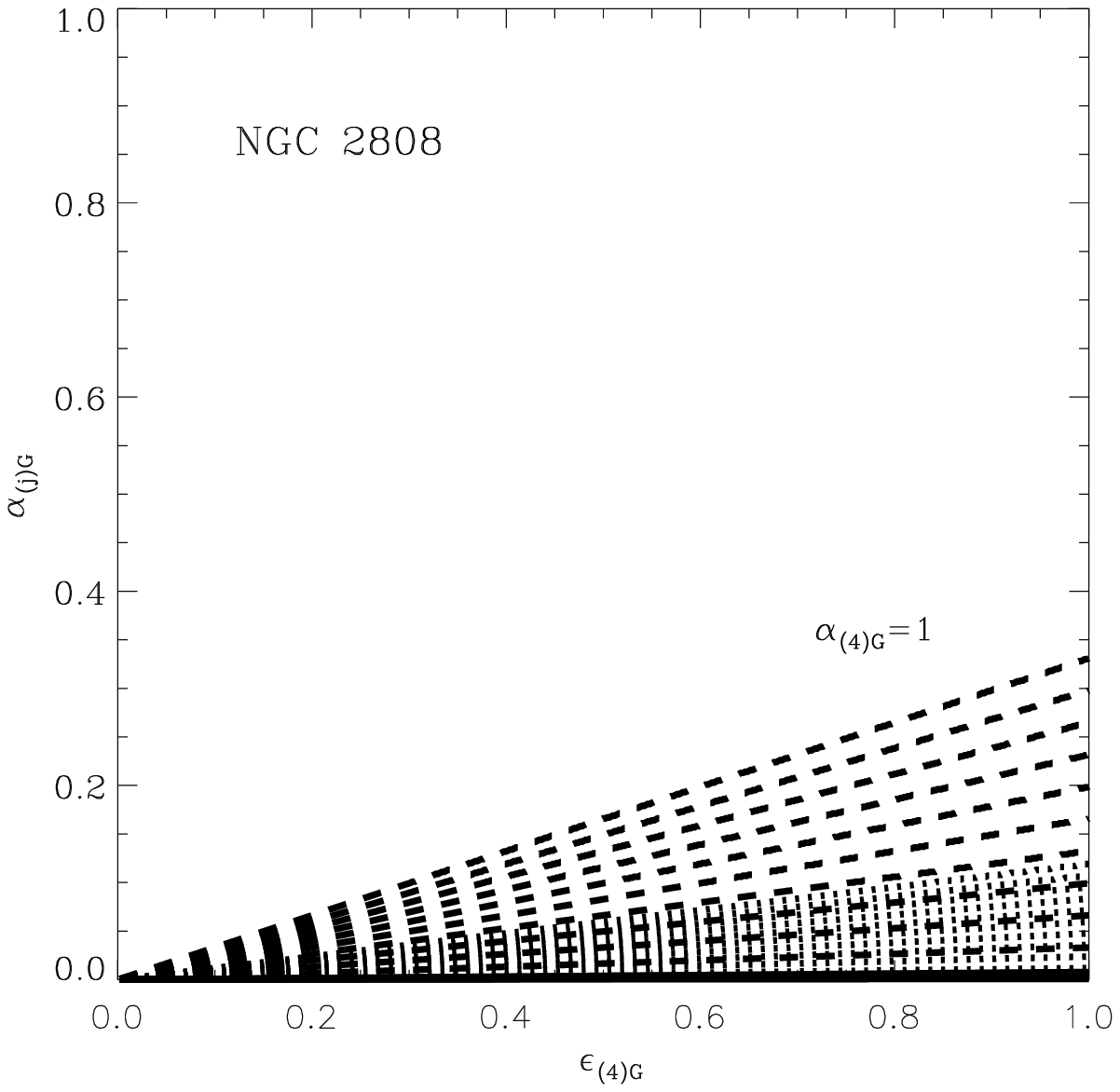}
\caption{Values of $\alpha_{(1)G}$ for our considered clusters, as a function of 
$\epsilon_{(n)G}$ and  a wide range of values of $\alpha_{(n)G}$. The upper two panels show
 $\alpha_{(1)G}$ for our two clusters with two stellar generations NGC 6266 and NGC 6397, and 
the lower panels show $\alpha_{(j)G}$ for NGC 6752 (a cluster with 3 stellar generations) and 
NGC 2808 that presents 4 stellar generations. In all panels the solid lines show the resultant values for 
$\alpha_{(1)G}$, dashed lines are for $\alpha_{(2)G}$ values and dotted lines, in the last panel 
give the $\alpha_{(3)G}$ of the third stellar generation. Note that in the last panel the solid lines appear almost at the x axes. Some of the dashed and dotted lines in  the 
lower panels do not reach the value of 1: $\alpha_{(2)G}$ (left panel) and $\alpha_{(3)G}$ (right panel),
as larger values would lead to $\alpha_{(1)G}$ (left panel) and $\alpha_{(2)G}$ (right panel) larger than 1.}
\label{fig2}
\end{figure}

The derived values of $\alpha_{(j)G}$  can be further used to convert present day
generation masses, $M_{(j)G}$, into the total original stellar mass of each generation ($M_{(j)*}$), accounting for their full IMF:
\begin{equation}
\label{eq17}
M_{(j)*} = M_{(j)G} / (\alpha_{(j)G} f_\mathrm{low})
\end{equation}
where $f_\mathrm{low}$ is a fraction of low mass stars that survive until
present in the stellar IMF. For standard IMF and a mass limit $0.8$\,M$_{\odot}$ it
is $f_\mathrm{low} \simeq 0.38$ \citep[e.g.][]{DeMink2009}.

Additionally, it is possible to use the star formation efficiencies,
$\epsilon_{(j)G}$, to convert the total  stellar generation masses ($M_{(j)*}$) to the mass of
the gas cloud out of which they were formed, $M_{(j)}$:
\begin{equation}
\label{eq18}
M_{(j)} = M_{(j)*} / \epsilon_{(j)G}
\end{equation}
where $\epsilon_{(j)G}$ is given by equation \ref{eq8}. Note that $M_{(1)} \equiv M_\mathrm{tot}$ 
is the mass of the original gaseous cloud out of which the whole cluster was formed.
Figure 4 displays the calculated  range of $M_{tot}$ for the considered clusters (see also Table 1).
This is shown as a function of the star formation efficiency of the last generation and for the wide range 
of values used for $\alpha_{(n)G}$. The procedure uses equations \ref{eq15} and \ref{eq16} to derive the 
$\alpha_{(j)G}$ values of each generation and equations \ref{eq17} and \ref{eq18} to derive the total 
mass in stars for each generation and the mass of the cloud out of which they formed to  finally obtain  
$M_\mathrm{tot}$.  Note that in all cases,  the lower limit to $M_\mathrm{tot}$ is the same for all 
allowed values of  $\alpha_{(n)G}$, while the upper limit is in all cases given by the lowest considered 
$\epsilon_{(n)G}$ and  
$\alpha_{(n)G}$ values. The upper limits for the mass of the primordial gas cloud result from demanding that the last generation of star formation, in each of the considered clusters, has been mostly lost
($\alpha_{(n)G}$ = 0.1) and thus the stars from the last generation that we see today represent only a small fraction of the stars initially formed. A small $\alpha_{(n)G}$ implies immediately a large    
$ M_\mathrm{tot}$, which makes one wonder why if 
$\epsilon_{(n)G}$ was also assumed to be so small, there was not another stellar generation. The Figure 
also shows that if all stars from the last generation were still  part of the cluster ($\alpha_{(n)G}$ = 1) the value of the 
 $ M_\mathrm{tot}$ upper limit would be reduced also by a factor of ten, for any assumed value of 
$\epsilon_{(n)G}$.
 
\begin{figure}[htbp]
\vspace{18.0cm}
\includegraphics{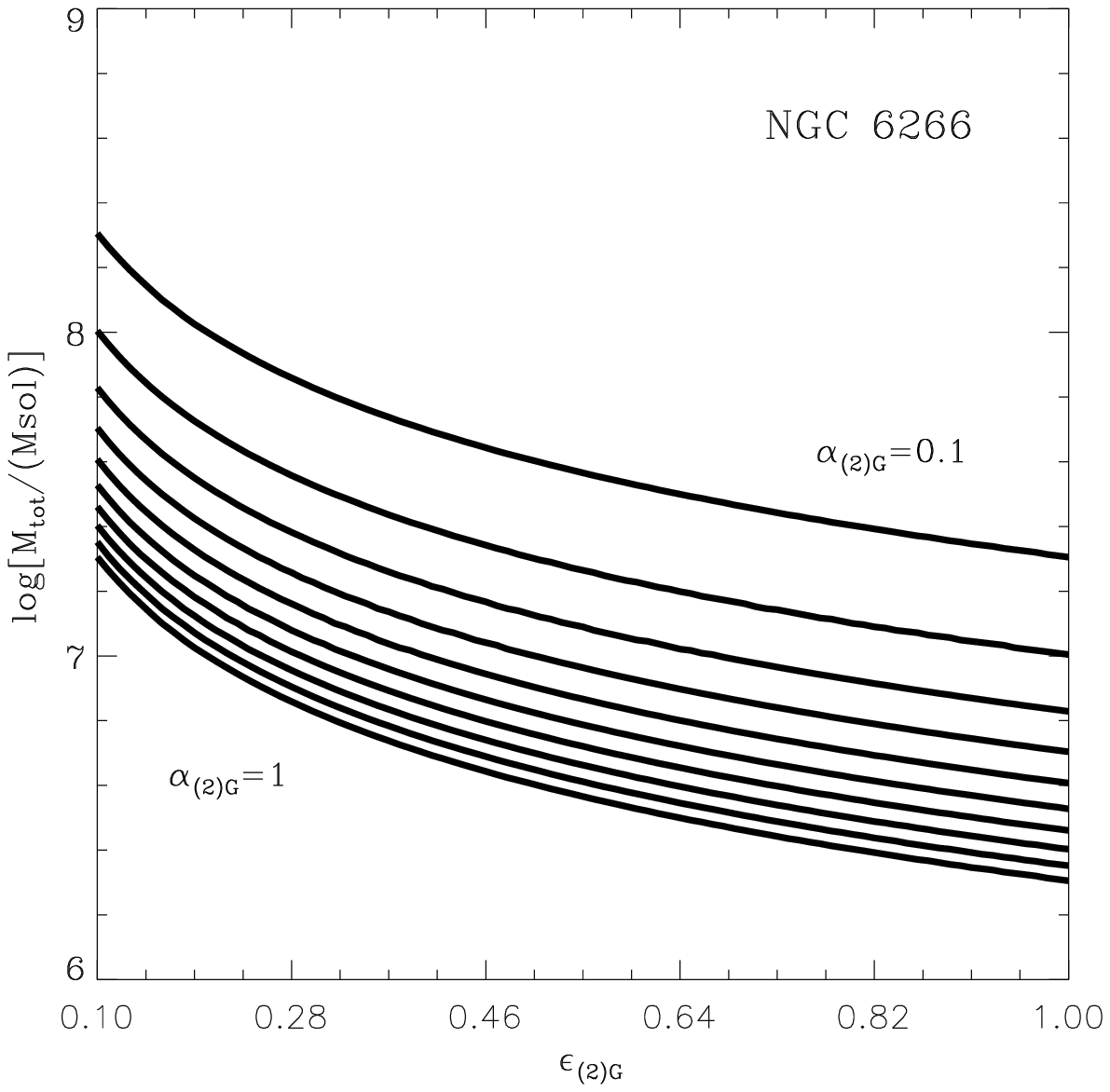}
\includegraphics{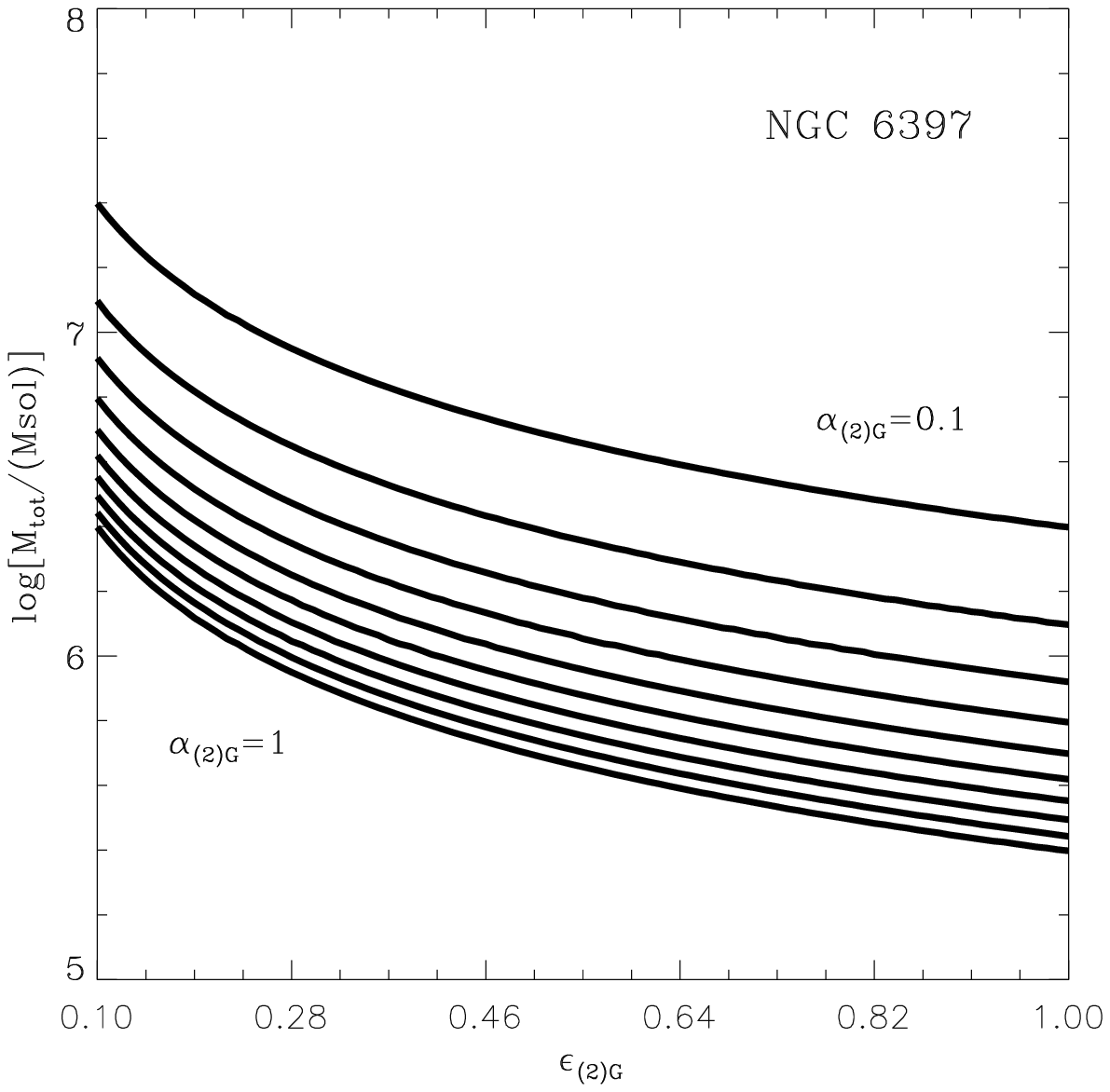}
\includegraphics{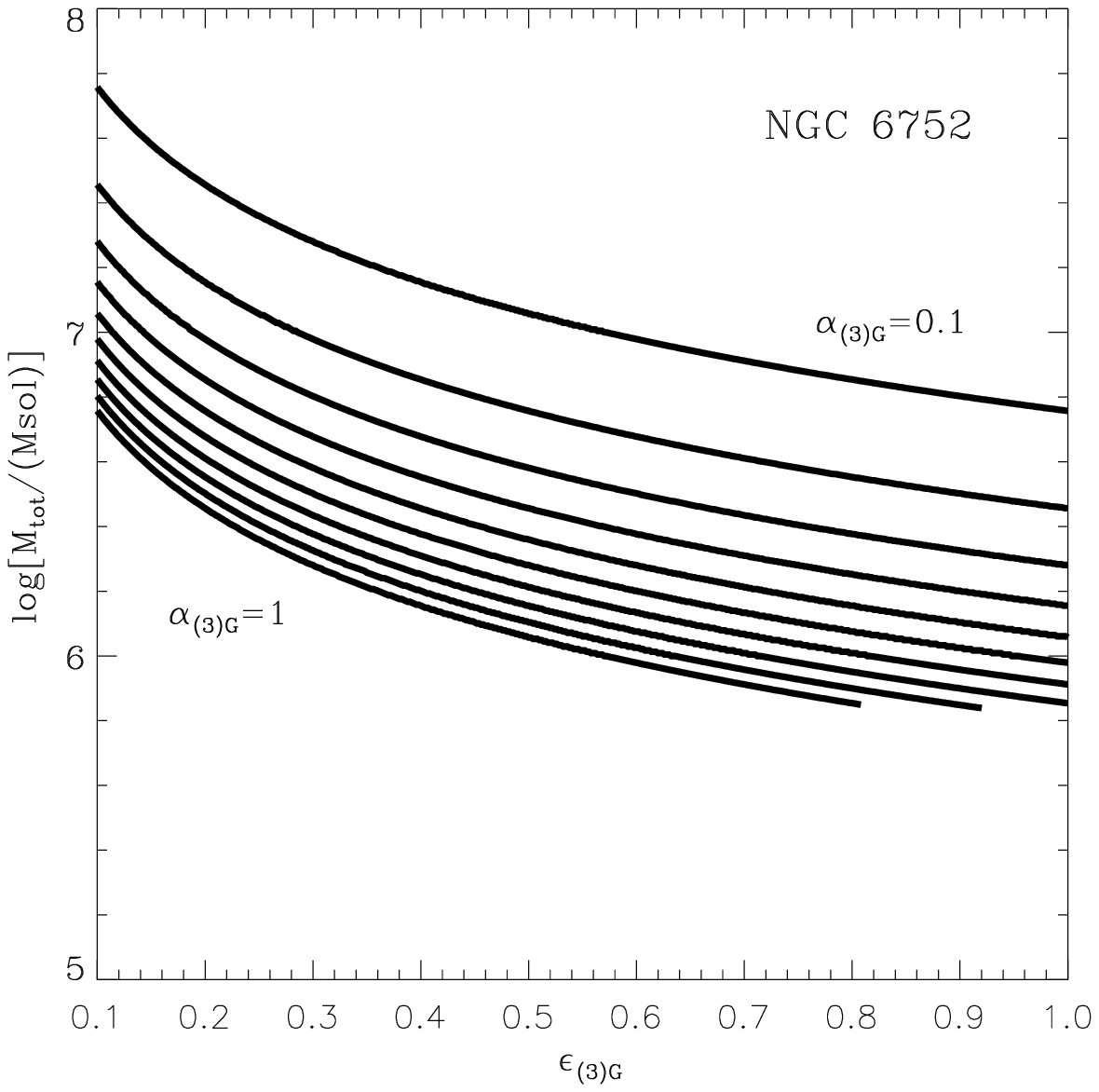}
\includegraphics{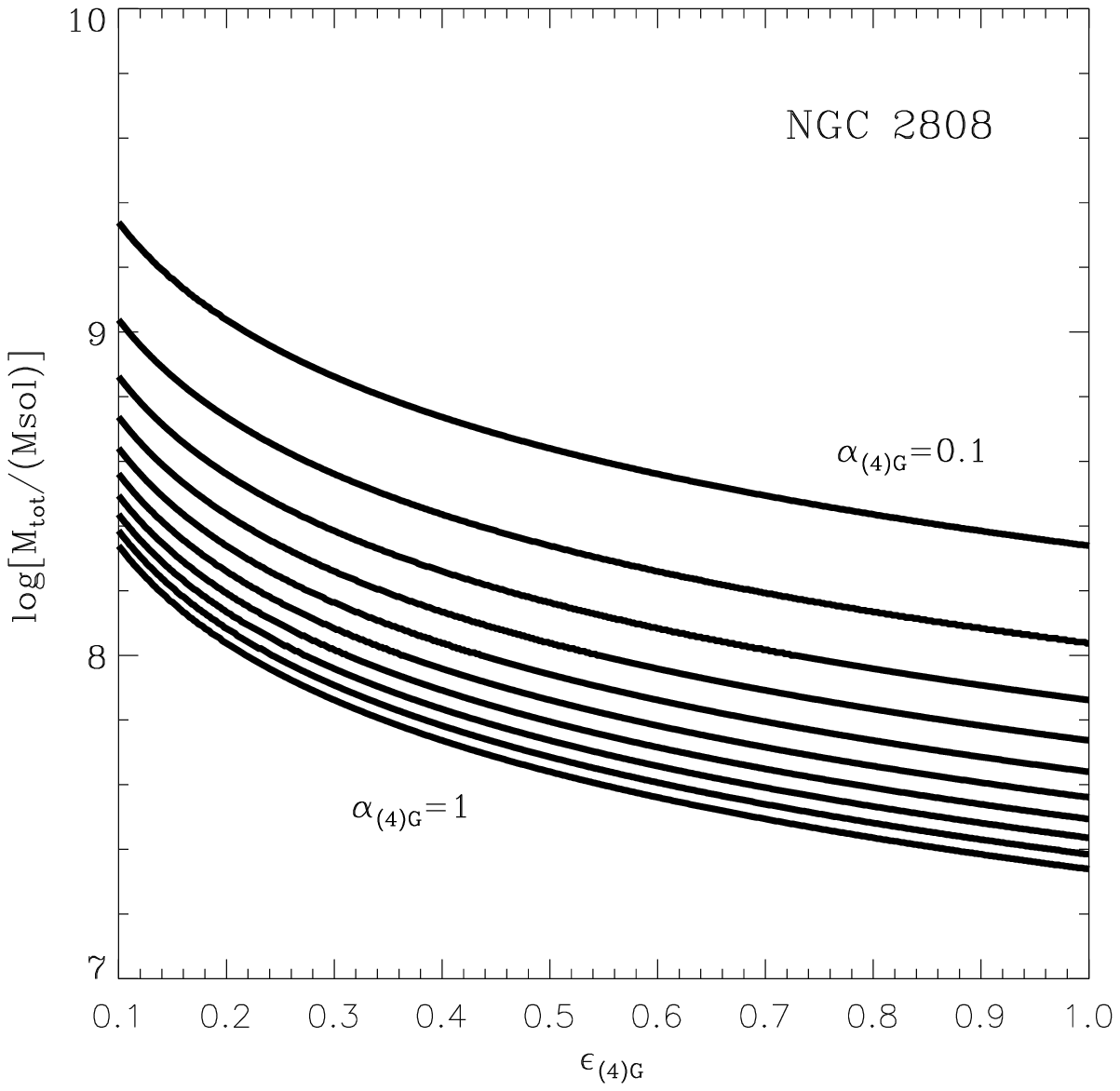}
\caption{The range  of  $M_\mathrm{tot}$ or $M_{1G}$ values derived for the clusters in our sample, 
plotted as a function of the assumed $\epsilon_{(n)G}$ and a wide range of  $\alpha_{(n)G}$ values.}
\label{fig3}
\end{figure}
   
\section{Concluding remarks}

Regarding the formation of GGCs we have shown that the efficiency of star formation in each stellar generation can be derived from  the helium enhancement $\Delta {Y}$ between consecutive generations. We have derived individual and total star formation efficiencies for a sample of GGCs. Regarding the evolution of clusters, we have shown that the relative fraction of stars from cosecutive  stellar generation is also directly dependent on the observed value of $\Delta {Y}$ between them. Finally, by means of some basic assumptions, and demanding for all generations to have originally a full IMF, we have shown how to infer  the original total mass in stars
for each stellar generation and have derived  the possible mass range of the clouds ($M_{tot}$) out of which each of our considered clusters formed.

\subsection{Evaluation of the model}

Here we use the main constraints imposed by photometric and spectroscopic studies on the models of 
secondary stellar generations in GGCs, derived by \citet{Renzini2015}, 
as a benchmark to evaluate our model. 

Variety and predominance are two important criteria to be taken into account for models of GGCs. Variety deals with the fact that no two GCs are identical. Some present a minimum of two stellar generations (as perhaps considered in most models in the literature) but other may present multiple (up to 7) generations. Predominance indicates that the present mass in the  various secondary generations exceeds in most cases the mass detected in the first generation. This issue leads in most models to a mass budget problem to account for the detected  contamination. This has led to assume a much larger first stellar generation,  now depleted by the loss of stars through tidal interactions and naturally to a much more massive primordial cloud. We assign an OK to our model in both issues, as done by \citet{Renzini2015}
for the original interactive massive binary model from \citet{DeMink2009}, 
noticing however that this model accounts only for two stellar generations and that the second one has a totally different IMF as it is to form  only  low mass stars.     

Color-magnitude diagrams (CMD) and appropriate two-colour plots indicate that star formation in GCs 
happened in bursts (called "discreteness" in \citet{Renzini2015}
and that it is not a continuous process, although see also the recent paper by 
\citet{Ventura2016} claiming instead a continuous process. In our model star formation with a full IMF 
occurs in well separated bursts. The collapse of the left over cloud is halted by the energy injected by the 
collection of massive stars that resulted from the last episode of star formation. As they evolve, 
they further contaminate the leftover gas with the new products that will characterize the next burst of star 
formation. The separation between bursts is then defined by the evolution time of massive stars ($\sim$ 
40-50 Myr).  This is in full agreement with  "discreteness" of stellar generations in GCs. 


The fact that in most GCs the various  generations of stars share the same [Fe/H] abundance imply a total 
inability of the residual clouds to  capture the ejecta from SN from former generations,  which 
\citet{Renzini2015} calls  SN avoidance. This in our model results from SN blowout, from the strong acceleration of the SN blast waves and of their shells of swept up matter, experienced either when they find the strong density 
gradient expected for the residual compact cloud held by gravity 
\citep[see][]{TenorioTagle2015,TenorioTagle2016}
or when overtaking several wind sources, which  by enhancing  the energy of the SN bubble makes them 
accelerate
\citep{Silich2017}. Acceleration leads to the immediate fragmentation of the swept up shell and to the rapid streaming of the SN ejecta between shell fragments and out of the remaining cloud. Only in the case of very massive and compact clouds SN taking place in the very central regions will be totally retained and thus in such massive cloud cases there is a possibility of enhancing the metal abundance of the left over cloud causing a noticible Fe spread, as in the most massive GCs.
   
In  our model, the observed helium enrichments $\Delta Y_{(j-1, j)}$  and masses $M_{(j)*}$ of
subsequent generations are taken into account defining  the star formation efficiencies and thus deserves an OK for both items. 
It is in fact He enrichment, the quantity  that defines here  the chronology of subsequent stellar generations as
well as their efficiency of star formation, which fully accounts for  the mass budget.

As in paper I,  interacting massive binaries are here held responsible of thorougly contaminating the leftover gas with H burning products while newly formed stars effectively hold the further collapse of the cloud. It is not until the end of the type II supernova (SN) era (say $\sim$ 40 Myr after the onset of the last episode of star formation) that gravity wins again and the collapse of the left over cloud leads to another stellar generation.

On the basis of the present analysis, we suggest the use of the $\Delta Y_{(j-1, j)}$ versus 
$\epsilon$  diagram as powerful tool for tracing the formation properties of Galactic GCs.

\section{Acknowledgments}

This work made use of v2.1 of the Binary Population and Spectral Synthesis (BPASS) models as last 
descrived by Eldridge, Stanway et al. 2017, Publications of the Astronomical Society of Australia 
(PASA 34, 61).

We thank our anonymous referee for many suggestions that greatly improved the clarity of the paper. 
This study was supported by CONACYT - M\'exico, research grants A1-S-28458 and by the 
Spanish Ministry of Science, Innovation and Universities the ESTALLIDOS grant AYA2016-79724-C4-2-P.
JP and RW acknowledge the support from project 19-15008S of the Czech Science Foundation and from 
the institutional project RVO:67985815.
\bibliographystyle{aasjournal}
\bibliography{GC}

\begin{thebibliography}{}
\expandafter\ifx\csname natexlab\endcsname\relax\def\natexlab#1{#1}\fi

\bibitem[{{Bastian} {et~al.}(2015){Bastian}, {Cabrera-Ziri}, \&
  {Salaris}}]{Bastian2015A}
{Bastian}, N., {Cabrera-Ziri}, I., \& {Salaris}, M. 2015, \mnras, 449, 3333

\bibitem[{{Bastian} {et~al.}(2013){Bastian}, {Lamers}, {de Mink}, {Longmore},
  {Goodwin}, \& {Gieles}}]{Bastian2013}
{Bastian}, N., {Lamers}, H.~J.~G.~L.~M., {de Mink}, S.~E., {et~al.} 2013,
  \mnras, 436, 2398

\bibitem[{{Bastian} \& {Lardo}(2018)}]{Bastian2018}
{Bastian}, N., \& {Lardo}, C. 2018, Annual Review of Astronomy and
  Astrophysics, 56, 83

\bibitem[{{Bedin} {et~al.}(2004){Bedin}, {Piotto}, {Anderson}, {Cassisi},
  {King}, {Momany}, \& {Carraro}}]{Bedin2004}
{Bedin}, L.~R., {Piotto}, G., {Anderson}, J., {et~al.} 2004, \apjl, 605, L125

\bibitem[{{Calura} {et~al.}(2015){Calura}, {Few}, {Romano}, \&
  {D'Ercole}}]{Calura2015}
{Calura}, F., {Few}, C.~G., {Romano}, D., \& {D'Ercole}, A. 2015, \apjl, 814,
  L14

\bibitem[{{Cassisi} {et~al.}(2017){Cassisi}, {Salaris}, {Pietrinferni}, \&
  {Hyder}}]{Cassisi2017}
{Cassisi}, S., {Salaris}, M., {Pietrinferni}, A., \& {Hyder}, D. 2017, \mnras,
  464, 2341

\bibitem[{{Chevalier} \& {Clegg}(1985)}]{Chevalier1985}
{Chevalier}, R.~A., \& {Clegg}, A.~W. 1985, \nat, 317, 44

\bibitem[{{D'Antona} {et~al.}(2002){D'Antona}, {Caloi}, {Montalb{\'a}n},
  {Ventura}, \& {Gratton}}]{DAntona2002}
{D'Antona}, F., {Caloi}, V., {Montalb{\'a}n}, J., {Ventura}, P., \& {Gratton},
  R. 2002, \aap, 395, 69

\bibitem[{{D'Antona} {et~al.}(2016){D'Antona}, {Vesperini}, {D'Ercole},
  {Ventura}, {Milone}, {Marino}, \& {Tailo}}]{DAntona2016}
{D'Antona}, F., {Vesperini}, E., {D'Ercole}, A., {et~al.} 2016, \mnras, 458,
  2122

\bibitem[{{de Mink} {et~al.}(2009){de Mink}, {Pols}, {Langer}, \&
  {Izzard}}]{DeMink2009}
{de Mink}, S.~E., {Pols}, O.~R., {Langer}, N., \& {Izzard}, R.~G. 2009, \aap,
  507, L1

\bibitem[{{Denissenkov} \& {Hartwick}(2014)}]{Denissenkov2014}
{Denissenkov}, P.~A., \& {Hartwick}, F.~D.~A. 2014, \mnras, 437, L21

\bibitem[{{Denissenkov} {et~al.}(2015){Denissenkov}, {VandenBerg}, {Hartwick},
  {Herwig}, {Weiss}, \& {Paxton}}]{Denissenkov2015}
{Denissenkov}, P.~A., {VandenBerg}, D.~A., {Hartwick}, F.~D.~A., {et~al.} 2015,
  \mnras, 448, 3314

\bibitem[{{D'Ercole} {et~al.}(2008){D'Ercole}, {Vesperini}, {D'Antona},
  {McMillan}, \& {Recchi}}]{DErcole2008}
{D'Ercole}, A., {Vesperini}, E., {D'Antona}, F., {McMillan}, S.~L.~W., \&
  {Recchi}, S. 2008, \mnras, 391, 825

\bibitem[{{Eldridge} {et~al.}(2017){Eldridge}, {Stanway}, {Xiao}, {McClelland},
  {Taylor}, {Ng}, {Greis}, \& {Bray}}]{Eldridge2017}
{Eldridge}, J.~J., {Stanway}, E.~R., {Xiao}, L., {et~al.} 2017, \pasa, 34, e058

\bibitem[{{Izzard} {et~al.}(2013){Izzard}, {de Mink}, {Pols}, {Langer}, {Sana},
  \& {de Koter}}]{Izzard2013}
{Izzard}, R.~G., {de Mink}, S.~E., {Pols}, O.~R., {et~al.} 2013, \memsai, 84,
  171

\bibitem[{{Kim} \& {Lee}(2018)}]{Kim2018}
{Kim}, J.~J., \& {Lee}, Y.-W. 2018, \apj, 869, 35

\bibitem[{{Krause} {et~al.}(2013){Krause}, {Charbonnel}, {Decressin}, {Meynet},
  \& {Prantzos}}]{Krause2013}
{Krause}, M., {Charbonnel}, C., {Decressin}, T., {Meynet}, G., \& {Prantzos},
  N. 2013, \aap, 552, A121

\bibitem[{{Krause} {et~al.}(2012){Krause}, {Charbonnel}, {Decressin}, {Meynet},
  {Prantzos}, \& {Diehl}}]{Krause2012}
{Krause}, M., {Charbonnel}, C., {Decressin}, T., {et~al.} 2012, \aap, 546, L5

\bibitem[{{Kroupa}(2001)}]{Kroupa2001A}
{Kroupa}, P. 2001, \mnras, 322, 231

\bibitem[{{Marino} {et~al.}(2014){Marino}, {Milone}, {Przybilla}, {Bergemann},
  {Lind}, {Asplund}, {Cassisi}, {Catelan}, {Casagrande}, {Valcarce}, {Bedin},
  {Cort{\'e}s}, {D'Antona}, {Jerjen}, {Piotto}, {Schlesinger}, {Zoccali}, \&
  {Angeloni}}]{Marino2014}
{Marino}, A.~F., {Milone}, A.~P., {Przybilla}, N., {et~al.} 2014, \mnras, 437,
  1609

\bibitem[{{Milone} {et~al.}(2012{\natexlab{a}}){Milone}, {Marino}, {Piotto},
  {Bedin}, {Anderson}, {Aparicio}, {Cassisi}, \& {Rich}}]{Milone2012C}
{Milone}, A.~P., {Marino}, A.~F., {Piotto}, G., {et~al.} 2012{\natexlab{a}},
  \apj, 745, 27

\bibitem[{{Milone} {et~al.}(2012{\natexlab{b}}){Milone}, {Piotto}, {Bedin},
  {King}, {Anderson}, {Marino}, {Bellini}, {Gratton}, {Renzini}, {Stetson},
  {Cassisi}, {Aparicio}, {Bragaglia}, {Carretta}, {D'Antona}, {Di Criscienzo},
  {Lucatello}, {Monelli}, \& {Pietrinferni}}]{Milone2012A}
{Milone}, A.~P., {Piotto}, G., {Bedin}, L.~R., {et~al.} 2012{\natexlab{b}},
  \apj, 744, 58

\bibitem[{{Milone} {et~al.}(2012{\natexlab{c}}){Milone}, {Piotto}, {Bedin},
  {Aparicio}, {Anderson}, {Sarajedini}, {Marino}, {Moretti}, {Davies},
  {Chaboyer}, {Dotter}, {Hempel}, {Mar{\'{\i}}n-Franch}, {Majewski}, {Paust},
  {Reid}, {Rosenberg}, \& {Siegel}}]{Milone2012B}
---. 2012{\natexlab{c}}, \aap, 540, A16

\bibitem[{{Milone} {et~al.}(2013){Milone}, {Marino}, {Piotto}, {Bedin},
  {Anderson}, {Aparicio}, {Bellini}, {Cassisi}, {D'Antona}, {Grundahl},
  {Monelli}, \& {Yong}}]{Milone2013}
{Milone}, A.~P., {Marino}, A.~F., {Piotto}, G., {et~al.} 2013, \apj, 767, 120

\bibitem[{{Milone} {et~al.}(2015{\natexlab{a}}){Milone}, {Marino}, {Piotto},
  {Bedin}, {Anderson}, {Renzini}, {King}, {Bellini}, {Brown}, {Cassisi},
  {D'Antona}, {Jerjen}, {Nardiello}, {Salaris}, {Marel}, {Vesperini}, {Yong},
  {Aparicio}, {Sarajedini}, \& {Zoccali}}]{Milone2015A}
---. 2015{\natexlab{a}}, \mnras, 447, 927

\bibitem[{{Milone} {et~al.}(2015{\natexlab{b}}){Milone}, {Marino}, {Piotto},
  {Renzini}, {Bedin}, {Anderson}, {Cassisi}, {D'Antona}, {Bellini}, {Jerjen},
  {Pietrinferni}, \& {Ventura}}]{Milone2015B}
---. 2015{\natexlab{b}}, \apj, 808, 51

\bibitem[{{Milone} {et~al.}(2017){Milone}, {Piotto}, {Renzini}, {Marino},
  {Bedin}, {Vesperini}, {D'Antona}, {Nardiello}, {Anderson}, {King}, {Yong},
  {Bellini}, {Aparicio}, {Barbuy}, {Brown}, {Cassisi}, {Ortolani}, {Salaris},
  {Sarajedini}, \& {van der Marel}}]{Milone2017}
{Milone}, A.~P., {Piotto}, G., {Renzini}, A., {et~al.} 2017, \mnras, 464, 3636

\bibitem[{{Milone} {et~al.}(2018){Milone}, {Marino}, {Renzini}, {D'Antona},
  {Anderson}, {Barbuy}, {Bedin}, {Bellini}, {Brown}, {Cassisi}, {Cordoni},
  {Lagioia}, {Nardiello}, {Ortolani}, {Piotto}, {Sarajedini}, {Tailo}, {van der
  Marel}, \& {Vesperini}}]{Milone2018}
{Milone}, A.~P., {Marino}, A.~F., {Renzini}, A., {et~al.} 2018, \mnras, 481,
  5098

\bibitem[{{Prantzos} \& {Charbonnel}(2006)}]{Prantzos2006}
{Prantzos}, N., \& {Charbonnel}, C. 2006, \aap, 458, 135

\bibitem[{{Renzini} {et~al.}(2015){Renzini}, {D'Antona}, {Cassisi}, {King},
  {Milone}, {Ventura}, {Anderson}, {Bedin}, {Bellini}, {Brown}, {Piotto}, {van
  der Marel}, {Barbuy}, {Dalessandro}, {Hidalgo}, {Marino}, {Ortolani},
  {Salaris}, \& {Sarajedini}}]{Renzini2015}
{Renzini}, A., {D'Antona}, F., {Cassisi}, S., {et~al.} 2015, \mnras, 454, 4197

\bibitem[{{Salaris} {et~al.}(2004){Salaris}, {Riello}, {Cassisi}, \&
  {Piotto}}]{Salaris2004}
{Salaris}, M., {Riello}, M., {Cassisi}, S., \& {Piotto}, G. 2004, \aap, 420,
  911

\bibitem[{{Sana} {et~al.}(2012){Sana}, {de Mink}, {de Koter}, {Langer},
  {Evans}, {Gieles}, {Gosset}, {Izzard}, {Le Bouquin}, \&
  {Schneider}}]{Sana2012}
{Sana}, H., {de Mink}, S.~E., {de Koter}, A., {et~al.} 2012, Science, 337, 444

\bibitem[{{Sbordone} {et~al.}(2011){Sbordone}, {Salaris}, {Weiss}, \&
  {Cassisi}}]{Sbordone2011}
{Sbordone}, L., {Salaris}, M., {Weiss}, A., \& {Cassisi}, S. 2011, \aap, 534,
  A9

\bibitem[{{Schneider} {et~al.}(2014){Schneider}, {Izzard}, {de Mink}, {Langer},
  {Stolte}, {de Koter}, {Gvaramadze}, {Hu{\ss}mann}, {Liermann}, \&
  {Sana}}]{Schneider2014}
{Schneider}, F.~R.~N., {Izzard}, R.~G., {de Mink}, S.~E., {et~al.} 2014, \apj,
  780, 117

\bibitem[{{Silich} \& {Tenorio-Tagle}(2017)}]{Silich2017}
{Silich}, S., \& {Tenorio-Tagle}, G. 2017, \mnras, 465, 1375

\bibitem[{{Silich} \& {Tenorio-Tagle}(2018)}]{Silich2018}
---. 2018, \mnras, 478, 5112

\bibitem[{{Sz{\'e}csi} \& {W{\"u}nsch}(2018)}]{Szecsi2018}
{Sz{\'e}csi}, D., \& {W{\"u}nsch}, R. 2018, ArXiv e-prints, arXiv:1809.01395

\bibitem[{{Tenorio-Tagle} {et~al.}(2016){Tenorio-Tagle},
  {Mu{\~n}oz-Tu{\~n}{\'o}n}, {Cassisi}, \& {Silich}}]{TenorioTagle2016}
{Tenorio-Tagle}, G., {Mu{\~n}oz-Tu{\~n}{\'o}n}, C., {Cassisi}, S., \& {Silich},
  S. 2016, \apj, 825, 118

\bibitem[{{Tenorio-Tagle} {et~al.}(2015){Tenorio-Tagle},
  {Mu{\~n}oz-Tu{\~n}{\'o}n}, {Silich}, \& {Cassisi}}]{TenorioTagle2015}
{Tenorio-Tagle}, G., {Mu{\~n}oz-Tu{\~n}{\'o}n}, C., {Silich}, S., \& {Cassisi},
  S. 2015, \apjl, 814, L8

\bibitem[{{Ventura} {et~al.}(2016){Ventura}, {Garc{\'{\i}}a-Hern{\'a}ndez},
  {Dell'Agli}, {D'Antona}, {M{\'e}sz{\'a}ros}, {Lucatello}, {Di Criscienzo},
  {Shetrone}, {Tailo}, {Tang}, \& {Zamora}}]{Ventura2016}
{Ventura}, P., {Garc{\'{\i}}a-Hern{\'a}ndez}, D.~A., {Dell'Agli}, F., {et~al.}
  2016, \apjl, 831, L17

\bibitem[{{Vink}(2018)}]{Vink2018}
{Vink}, J.~S. 2018, \aap, 615, A119

\end{thebibliography}

\end{document}